\UseRawInputEncoding
\documentclass[twocolumn,showpacs,preprintnumbers,amsmath,amssymb,superscriptaddress,aps]{revtex4}

\usepackage{graphicx}
\usepackage{dcolumn}
\usepackage{bm}
\usepackage{amsfonts}
\usepackage{mathrsfs}
\usepackage{graphicx}                   
\usepackage{dcolumn}
\usepackage{bm}
\usepackage{amsmath}

\begin{document}


\title{Giant electro-optic effect in chiral topological semimetal RhSi}

\author{Zhi Li}
\email[Email:]{zhili@njust.edu.cn}
\affiliation{MIIT key Laboratory of Advanced Display Materials and Devices, Ministry of Industry and Information Technology, Institute of optoelectronics $\&$ Nanomaterials, Nanjing University of Science and Technology, Nanjing, 210094, China}

\author{Yuewen Gao}
\affiliation{MIIT key Laboratory of Advanced Display Materials and Devices, Ministry of Industry and Information Technology, Institute of optoelectronics $\&$ Nanomaterials, Nanjing University of Science and Technology, Nanjing, 210094, China}
\author{Yu Gu}
\affiliation{MIIT key Laboratory of Advanced Display Materials and Devices, Ministry of Industry and Information Technology, Institute of optoelectronics $\&$ Nanomaterials, Nanjing University of Science and Technology, Nanjing, 210094, China}
\author{Shengli Zhang}
\affiliation{MIIT key Laboratory of Advanced Display Materials and Devices, Ministry of Industry and Information Technology, Institute of optoelectronics $\&$ Nanomaterials, Nanjing University of Science and Technology, Nanjing, 210094, China}

\author{Toshiaki Iitaka}
\affiliation{Discrete Event Simulation Research Team, RIKEN Center for Computational Science, 2-1 Hirosawa, Wako, Saitama, 351-0198 Japan}

\author{W. M. Liu}
\email[Email:]{wmliu@iphy.ac.cn}
\affiliation{Beijing National Laboratory for Condensed Matter Physics, and Institute of Physics, Chinese Academy of Sciences, Beijing 100190, China}

\date{\today}

\begin{abstract}
 We studied the linear electro-optic effect of chiral topological semimetal RhSi which is characterized by high-fold
chiral fermions separated in energy space. We identify that the general second order conductivity $\sigma^{(2)}_{xyz}(\omega=\omega_{1}+\omega_{2};\omega_{1},\omega_{2})$ includes a real symmetric component and an imaginary antisymmetric component, which are from the inter-band shift and intra-band injection current with frequency $\omega$, respectively. The $\sigma^{(2)}_{xyz}$ is significantly enhanced by the high electron velocity and nontrivial band topology of chiral fermion, and modifies the phase velocity of light wave. We also predict that the electro-optic coefficient $\chi_{xyz}^{(2)}(\omega;\omega,0)$ of chiral crystal RhSi is about 7000 pm/V at photon energy 0.01 eV and 1.1 eV, which is about 200 times that of widely used LiNbO$_{3}$ crystal. The giant electro-optic coefficient renders a relatively low half-wave voltage in order of hundreds volt, and demonstrates potential application as electro-optic crystals for the wavelength in the second telecom window of optical fiber communications.
\end{abstract}

\maketitle

    Electro-optic modulator is one fundamental active device in the optical interconnect systems \cite{EK21}. However, silicon doesn't have linear electro-optic effect as the compound semiconductor LiNbO$_{3}$ which has relatively large electro-optic coefficient \cite{Boyd,QL20,ML18}. The electro-optic coefficient of commercial LiNbO$_{3}$ crystal is 32.6 pm/V at 500 nm. Because of this relativity weak electro-optic coefficient, the half wave-voltage is in order of 10 kV for visible light \cite{Boyd}, which is much higher than the drive voltage of CMOS circuit. Usually, silicon-based modulators make use of plasma dispersion effect \cite{SB87,MP04,ML05,XX18,A17}, instead of the linear electro-optic effect. Because of tunable Fermi level under bias voltage and high carrier mobility, graphene-based waveguide integrated electro-absorption modulator \cite{XZ11,ML15,KH19} has advantages of broad optical bandwidth and high operation speed. Since both spatial inversion and time-reversal invariant symmetry are preserved in pristine graphene, linear electro-optic effect is vanishing and phase modulation is impossible. However, the fourfold band degeneracy of Dirac electron will split into a pair of massless chiral fermions with opposite chirality, if either spatial inversion or time-reversal invariant symmetry is broken. The intrinsic mechanism for linear electro-optic effect in topological matters with chiral fermion is an uncharted topic before.

   Transition metal silicides CoSi, RhSi, and PdGa crystalize into simple cubic structure with space group $P2_{1}3$ \cite{RhSi} which includes no spatial inversion symmetry or mirror symmetry, and they afford an excellent platform for the realization of high-fold chiral fermions \cite{DH21,AV18,JO21}. In the band structure of RhSi by calculation and angle-resolved photoelectron spectroscopy (ARPES) experiments \cite{MZH18,Sato19,MZH17,ZSC17}, the spin-1 fermion at $\Gamma$ point and topological charge-2 fermion at R point are separated in energy scale because of absent mirror symmetry. Absolute magnitude of the Chern number with the Berry curvature around $\Gamma$ point is determined to 2 in CoSi and RhSi by ARPES measurements \cite{DH19,MZH19,YL19}. However, higher Chern number C=4 is observed in PdGa which has much stronger spin-orbital coupling SOC \cite{CF20,AVP21}. Except the novel high-fold chiral fermions in chiral topological semimetal RhSi, photogalvanic effect are also widely studied for potential application in field of infrared or Terahertz detection, and possible quantized circular photogalvanic effect \cite{JO21}. The first-principles calculations predicted relatively strong second order optical conductivity $\sigma_{xyz}^{(2)}(0;\omega,-\omega)=\sigma_{xyz}+\eta_{xyz}$ in topological semimetals RhSn and RhSi in range of Terahertz \cite{ZL21B,ZL19}. By effective model Hamiltonian and the first-principles calculations, circular photogalvanic effect is predicted in chiral topological semimetal RhSi \cite{CPGE1,NN18,TM18,YS20}. Experimentally, the circular photogalvanic effect in the chiral topological semimetal RhSi is observed in an energy window below 0.65 eV \cite{JO19,JE20,DHT21}. The linear electro-optic effect, is essentially a second order response under high frequency optical signal and a low frequency electric signal. The chiral topological semimetal also offers an excellent platform to investigate the intrinsic mechanism for linear electro-optic effect from the high-fold chiral fermions with opposite chirality separated in energy space.

    In this Letter, we studied the linear electro-optic effect in chiral topological semimetal RhSi. We identify that the the general complex optical conductivity $\sigma_{ijk}^{(2)}(\omega;\omega_{1},\omega_{2})$ should also include a symmetric real component from inter-band contribution and an antisymmetric imaginary component from intra-band contribution. We calculate the electro-optic susceptibility of RhSi by the first-principles calculation within sum-over-state approximation, and predict a giant electro-optic coefficient 7000 pm/V. This result shows that chiral topological semimetal RhSi has potential application in phase modulation with relatively low half-wave voltage for wavelength in the second telecom window of optical fiber communications.


    With respect to photogalvanic effect, we focus on the dc photocurrent. However, for linear electro-optic effect, the ac current with the same frequency of optical signal is important. The ac current should include oscillating currents from both linear and second order response. Within the dipole approximation \cite{TGP17}, the total Hamiltonian describes the light-matter interaction reads $H=h_{0}-e\vec{E}\cdot \vec{r}$, in which $h_{0}$ describes the ground state and $\vec{E}$ is the electric field. The velocity operator reads $\vec{v}(t)=\frac{i}{\hbar}[H,\vec{r}(t)]$ in which position operator $\vec{r}(t)=e^{-iHt/\hbar}\vec{r}e^{-iHt/\hbar}$. The current density reads $\vec{j}(t)=e\langle \psi (\vec{r},t)|\vec{v}(t)|\psi (\vec{r},t)\rangle$, in which the field operator $\psi (r,t)$ can be expanded by Bloch functions. By some lengthy but straightforward calculations, the second order response contributed oscillating current $\vec{j}^{(2)}(\omega)$ reads,
\begin{equation}\label{current}
  j_{i}^{(2)}(\omega)=\sum_{\omega_{1},\omega_{2}}\sigma_{ijk}^{(2)}(\omega;\omega_{1},\omega_{2})E_{j}(\omega_{1})E_{k}(\omega_{2})\delta(\omega,\omega_{1}+\omega_{2}),
\end{equation}
  which frequency dependent second order conductivity $\sigma_{ijk}^{(2)}(\omega;\omega_{1},\omega_{2})=\alpha_{ijk}(\omega;\omega_{1},\omega_{2})+i\beta_{ijk}(\omega;\omega_{1},\omega_{2})$.
  The first term $\alpha_{ijk}(\omega;\omega_{1},\omega_{2})$ from the inter-band shift current is symmetric under the exchange of \emph{j} and \emph{k}, while the second term $\beta_{ijk}(\omega;\omega_{1},\omega_{2})$ from the intra-band injection current is antisymmetric under the exchange of \emph{j} and \emph{k}. However, shift and injection currents mentioned here are oscillating with frequency of optical signal, instead of dc photocurrent. In fact, $\sigma_{ijk}(\omega;\omega_{1},\omega_{2})$ is proportional the product of velocity $v^{i}$ and quantum geometric tensor $Q^{jk}=\Gamma^{jk}+i\Omega^{jk}$ which includes a symmetric metric tensor $\Gamma^{jk}$ and an antisymmetric curvature tensor $\Omega^{jk}$ \cite{WML10,AP17,FP21,DC21,NN20}. The quantum geometric tensor is responsible for the generation of carriers \cite{ZL211}. If we constrain $\omega$=0, Eq.~(1) describes the photogalvanic effect. With $\omega_{1}=\omega$, and $\omega_{2}$=0, Eq.~(1) describes the linear electro-optic effect. The detailed derivation of Eq.~(1) by density matrix method is present in the supplementary material \cite{SM}.

  Both components $\alpha_{ijk}(\omega;\omega_{1},\omega_{2})$ and $\beta_{ijk}(\omega;\omega_{1},\omega_{2})$ are related to the difference of electron velocities and energy between the two bands which are excited by external light field. With linear band dispersion, electron velocity is reduced to the Fermi velocity. Usually, the linear band dispersion in topological semimetals render a relatively high electron velocity and a low effective mass, which will enhance the nonlinear optical conductivity significantly. Especially, the nonlinear currents contributed by the chiral fermions at $\Gamma$ and $R$ points with opposite chirality will not cancel out each other because of they can not be resonantly excited simultaneously, leading to non-vanishing $\beta_{ijk}(\omega;\omega_{1},\omega_{2})$ which is related to Berry curvature. Under optical field with low photon energy, the Berry curvature around the $\Gamma$ point will dominate the injection current. Thus, higher Chern number will enhance the $\beta_{ijk}(\omega;\omega_{1},\omega_{2})$ at low energy. Additionally, the chiral topological semimetal RhSi has space group $P2_{1}3$ ($\#198$), and the only non-vanishing elements of second-order conductivity $\sigma_{ijk}^{(2)}(\omega;\omega_{1},\omega_{2})$ have indices \emph{xyz} and its permutations. Therefore, the ac current from second order contribution is always transverse to the plane spanned by optical field and electric field, i.e., ac nonlinear Hall current will present in chiral topological semimetal RhSi. Since chiral crystal RhSi has large second order optical conductivity $\sigma_{xyz}^{(2)}(0;\omega,-\omega)$ because high electron velocity, it may also has large electro-optic susceptibility, which connects the second order optical conductivity by $\sigma_{xyz}^{(2)}(\omega;\omega,0)=-i\omega\epsilon_{0}\chi^{(2)}_{xyz}(\omega;\omega,0)$ and $\epsilon_{0}$ is the vacuum electric permittivity.

  \begin{figure}[t]
  \includegraphics[width=8.5cm]{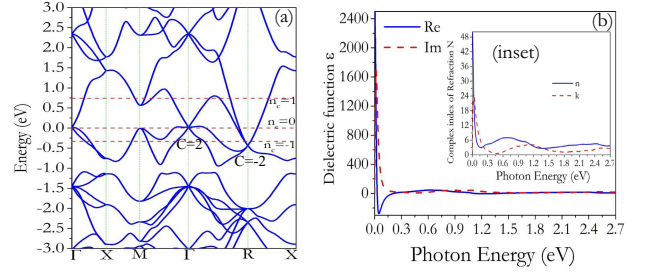}
  \caption{\label{fig1} (a) The band structure of RhSi crystal without SOC, where the doping concentration $n_{c}$ dependent Fermi level is marked by red dash line. Without doping, the Fermi level is set to 0 eV. The chiral fermions at $\Gamma$ on Fermi level and the $R$ point below Fermi level have opposite chirality. With electron (hole) doping, the Fermi level shifts to the higher (lower) red dash line. (b) Frequency dependent dielectric function $\epsilon(\omega)$ and complex refractive index $N=n(\omega)+ik(\omega)$ (inset) by the first-principles calculation within linear response. The peak at low frequency is dominated by intra-band contribution (Drude term).}
\end{figure}

   So far, we have clarified the characters of second-order conductivity for chiral topological semimetal RhSi. We should further clarify how the nonlinear Hall current changes the propagation of optical signal in RhSi crystal. We consider a light beam with propagating vector $q$ along \emph{z}-direction and linear polarization in \emph{xy} plane, and an applied (static or low frequency) bias voltage along \emph{z}-direction. For monochromatic light with frequency of $\omega$, the total electron current oscillating with frequency of $\omega$ reads $j_{i}(\omega)=\sigma_{ij}(\omega)E_{j}(\omega)+j_{i}^{(2)}(\omega)$, in which the first term is contributed by the linear response including both intra-band and inter-band contributions. From Eq. ~(1), the nonlinear Hall current with frequency $\omega$ reads
  \begin{eqnarray}
    j^{(2)}_{x}(\omega) &=& \sigma_{xyz}^{(2)}(\omega;\omega,0)E_{z}E_{y}(\omega),\\
    j^{(2)}_{y}(\omega) &=& \sigma_{xyz}^{(2)*}(\omega;\omega,0)E_{z}E_{x}(\omega).
  \end{eqnarray}
  From the macroscopic equations of Maxwell's equations, the complex refractive index $\tilde{N}(\omega)=\frac{cq}{\omega}$ (\emph{c} the speed of light in vacuum) satisfies,
  \begin{equation}\label{MW}
    \left(
       \begin{array}{cc}
        \tilde{N}^{2}-N^{2}(\omega)  & \sigma_{xyz}^{(2)}(\omega;\omega,0)E_{z} \\
        \sigma_{xyz}^{(2)*}(\omega;\omega,0)E_{z} & \tilde{N}^{2}-N^{2}(\omega) \\
       \end{array}
     \right
    )\left(
        \begin{array}{c}
          E_{x}(\omega) \\
          E_{y}(\omega) \\
        \end{array}
      \right
    )=0,
  \end{equation}
  in which  $N^{2}(\omega)=\epsilon(\omega)$ and $\epsilon(\omega)$ is the complex dielectric tensor from linear response theory.
  The two solutions for $\tilde{N}(\omega)$ read,
  \begin{equation}\label{Re}
    \tilde{N}_{\pm}(\omega)=\sqrt{N^{2}(\omega)\pm |\chi_{xyz}^{(2)}(\omega;\omega,0)|E_{z}},
  \end{equation}
  and $\sigma_{xyz}^{(2)}(\omega;\omega,0)=-i\omega \epsilon_{0} \chi^{(2)}_{xyz}(\omega;\omega,0)$. The two eigenvectors $\frac{1}{\sqrt{2}}|E(\omega)|(1,\pm 1)$ can be interpreted as two orthogonal optical fields with different polarizations, and $|E(\omega)|e^{-\omega\tilde{N}z/c}$ is the strength of propagating electromagnetic wave. Eq.~(7) also demonstrates that the two orthogonal optical fields in chiral crystal have different refractivities $\tilde{n}_{\pm}(\omega)=Re \tilde{N}_{\pm}(\omega)$. From Eq.~(7), the difference between $\tilde{n}_{+}(\omega)$ and $\tilde{n}_{-}(\omega)$ reads,
  \begin{equation}\label{pd}
    \Delta n(\omega)=\tilde{n}_{+}(\omega)-\tilde{n}_{-}(\omega)\sim\frac{|\chi^{(2)}_{xyz}(\omega;\omega,0)|E_{z}}{2n(\omega)},
  \end{equation}
 in which $n(\omega)$ is the real part of complex index of refraction $N(\omega)$ from linear response theory. After light beam passes through the electro-optic crystal, the phase difference between the two orthogonal optical fields is given by,
  \begin{equation}\label{Pd}
    \Delta\varphi=\Delta n(\omega)\frac{\omega L}{2c},
  \end{equation}
  in which L is the length of crystal along z-direction which is parallel to the propagating vector of light field. The half-wave voltage is given by,
  \begin{equation}\label{HWV}
    V_{\lambda/2}=\frac{2\pi c}{\omega |\chi^{(2)}_{xyz}(\omega;\omega,0)|}.
  \end{equation}

  To determine the frequency dependent half-wave voltage, we need the information of refractivity $n(\omega)$ and $\chi^{(2)}_{xyz}(\omega;\omega,0)$. Both refractivity $n(\omega)$ and $\chi^{(2)}_{xyz}(\omega;\omega,0)$ can be calculated by the first-principles calculation based on density functional theory. In this work, the linear optic properties of RhSi are calculated by all-electron full-potential linearised augmented-plane wave (LAPW) code ELK \cite{ELK}, while the second-order susceptibility  $\chi^{(2)}_{xyz}(\omega;\omega,0)$ is calculated by the Abinit code with sum-over-state approximation \cite{ABINIT,SOS}. At last, the frequency dependent half-wave voltage is calculated by Eq.~(8).

\begin{figure}[t]
  \includegraphics[width=8.5cm]{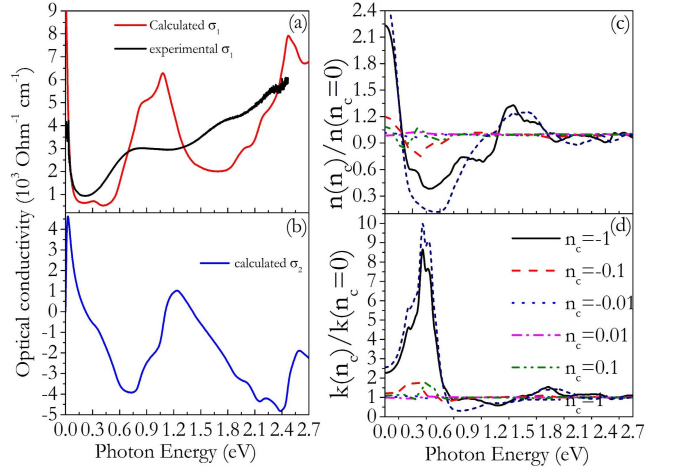}
  \caption{\label{fig2} Frequency and doping concentration dependent optical conductivity $\sigma(\omega)=\sigma_{1}+i\sigma_{2}$ including both intra-band and inter-band contributions by the first-principles calculation within linear response. (a) Calculated and experimental $\sigma_{1}$ from reference \cite{AVP20}; (b) Calculated $\sigma_{2}$ ; (c,d) Doping concentration dependent $n(n_{c})$/$n(n_{c}=0)$ and $k(n_{c})$/$k(n_{c}=0)$ by the first-principles calculation, which present relatively large change when doping concentration is larger than 0.1 electron or hole per unit cell.}
\end{figure}

  Since the SOC effect is weak in chiral topological semimental RhSi, we ignore the SOC effect in the first-principles calculation and adopt fine k-point grids 50$\times$50$\times$50 for linear optic properties with intra-band contribution (Drude term). The calculated band structure is shown in Fig.~1a, in which triply degenerated and doubly degenerated band crossings are reproduced at $\Gamma$ and $R$ points, respectively. Without doping, the Fermi level is set to zero energy. With doping concentration of one electron (hole) per unit cell, the Fermi level shifts to the position of higher(lower) red dash line. Without doping, the calculated plasma frequency is 1.39 eV, and the calculated dielectric function $\epsilon(\omega)$ is shown in Fig.~1b. The complex index of refraction N=$n(\omega)$+i$k(\omega)$ is shown in the inset of Fig.~1b. In the range of low frequency, the real parts of $\epsilon(\omega)$ and N are very high because of the dominating intra-band contribution. The calculated refractivity $n(\omega)$ is about 51, which is very close to the experimental result \cite{JO19}. Additionally, the extinction coefficient (or attenuation index) $k(\omega)$ is also very high in range of low frequency, indicating high absorption and rapid attenuation when the optical field propagates through the chiral RhSi crystal. The calculated real part of optic conductivity as shown in Fig. 2a is similar to the experimental result by Maulana et al. \cite{AVP20} and Ni et al. \cite{WL20}, except for one more sharp peak at low frequency, which may be contributed by ion conductivity. At photon energy 0.6-1.2 eV, the calculated result is much higher than the experimental result by Maulana et al., and the later is also smaller than the experimental result by Ni et al. The imaginary part of optic conductivity is shown in Fig. 2b. All off-diagonal elements of dielectric tensor and complex index of refraction are vanishing by linear response. Our calculated linear optic properties are consistent with experimental results very well, showing that the SOC is not important in chiral topological semimetal RhSi, which is also confirmed by ARPES experiments \cite{DH19,MZH19,YL19}. Overall, bulk RhSi has much high optical absorption, in contrast to pristine graphene monolayer.

  For graphene based electro-absorption modulator, the tunable Fermi level under electric field modulates the optic absorption. Similarly, the electric signal can also tune the Fermi level and the linear optic properties of RhSi. We calculate the doping concentration $n_{c}$ dependent dielectric function. Since we are interested with the changes of refractivity and extinction coefficients, we plot $n(n_{c})$/$n(n_{c}=0)$ and $k(n_{c})$/$k(n_{c}=0)$ in Fig.~2c and Fig.~2d, respectively. Here, positive (negative) $n_{c}$ means electron (hole) doping. The calculated results show that $n(n_{c})$/$n(n_{c}=0)$ and $k(n_{c})$/$k(n_{c}=0)$ will present significant change when doping concentration is larger than 0.1 electron or hole per unit cell, viz. 9.8$\times$ 10$^{20}$ /cm$^{3}$.

 \begin{figure}[t]
  \includegraphics[width=8.5cm]{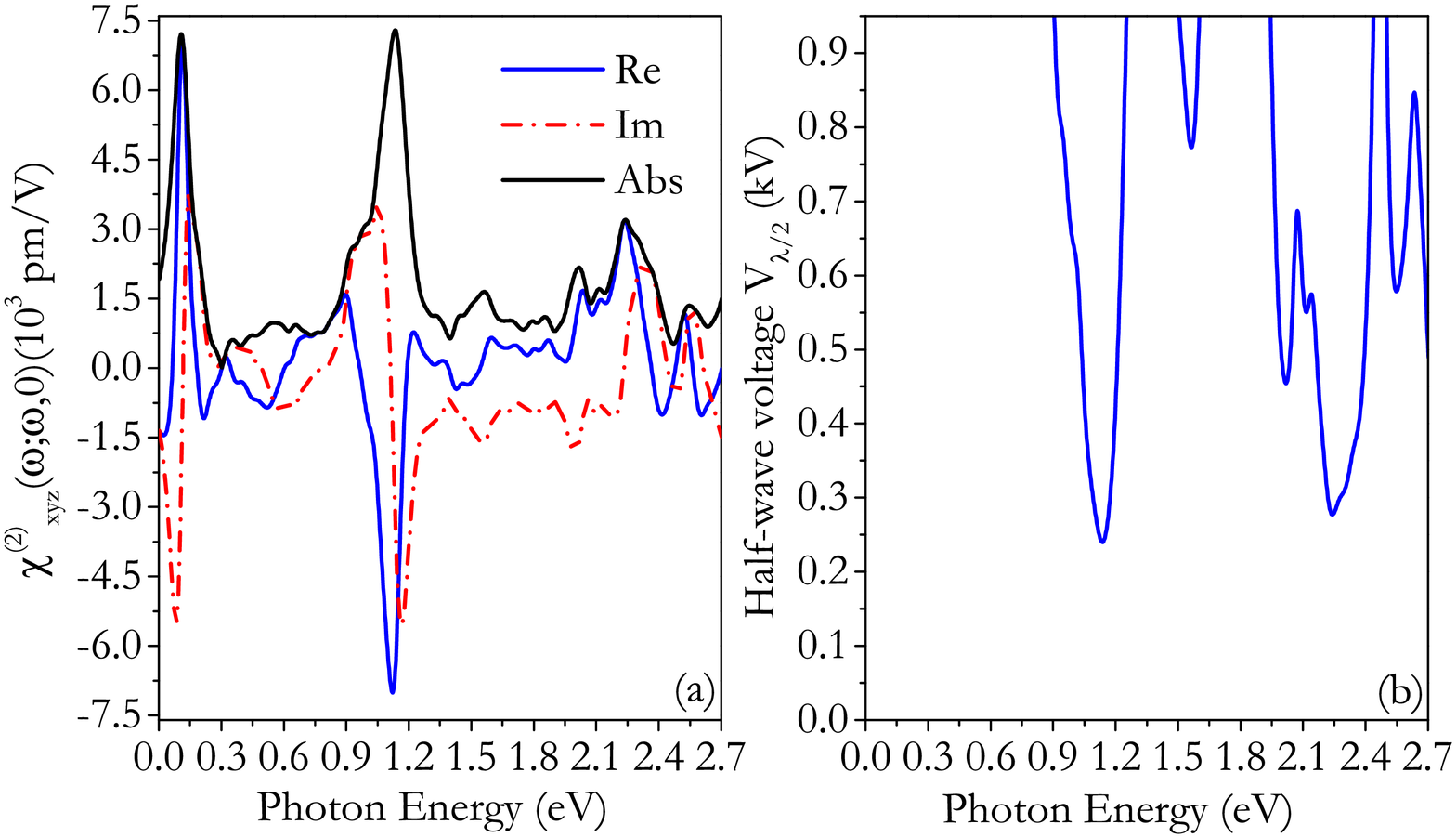}
  \caption{\label{fig3} (a) Frequency dependent electro-optic coefficient  $\chi^{(2)}_{xyz}(\omega;\omega,0)$, where the real and imaginary parts are plotted by blue real line and red dash dot line, respectively. The real part presents different sign at 0.01 eV and 1.1 eV. (b) Half-wave voltage determined by Eq. (8). At low frequency, the half-wave voltage is high and out of range.}
\end{figure}

  Since chiral crystal RhSi is nonmagnetic, time-reversal invariant symmetry is preserved. Therefore, both contributions from single band, including second order Ohm conductivity and Berry curvature dipole, are vanishing \cite{Fu15}. Within the sum-over-state approximation, the calculated frequency-dependent electro-optical susceptibility tensor $\chi^{(2)}_{xyz}(\omega;\omega,0)$ of RhSi is shown in Fig.~3a. At photon energy of 0.01 eV and 1.1 eV, the absolute value of electro-optical coefficient is as large as 7000 pm/V, which is much larger than that in widely used electro-optic crystal LiNbO$_{3}$ \cite{Boyd}. Since  $\sigma_{xyz}^{(2)}(\omega;\omega,0)=-i\omega\epsilon_{0}\chi^{(2)}_{xyz}(\omega;\omega,0)$, the real and imaginary part of $\chi^{(2)}_{xyz}(\omega;\omega,0)$ is related to the Berry curvature and metric tensor, respectively. The peak in the real part of electro-optical susceptibility at low photon energy 0.01 eV is resulting the chiral fermion at $\Gamma$ point. Since the injection currents contributed by chiral fermions at $\Gamma$ point and $R$ point have opposite sign, the other chiral fermion at $R$ point with opposition chirality will cancel the contribution from the chiral fermion at $\Gamma$ point with increasing photon energy. This scenario is similar to the circular photogalvanic effect observed below 0.65 eV \cite{JO19}. The other peak in the real part at 1.1 eV is related to the chiral fermion with lower energy below the Fermi level. We also note that the optical conductivity from linear response also presents peaks at 0.01 eV and 1.1 eV. It suggests that the electric signal plays the role of bias voltage which drives the oscillating carriers excited by optical signal to form transverse current with frequency the same as the optical signal. Since the real refractivity \emph{n} and imaginary part \emph{k} determines the phase velocity of light wave propagating in medium, giant real and imaginary parts of $\chi^{(2)}_{xyz}(\omega;\omega,0)$ mean relatively rapid changes of phase velocity modulated by electric signal. We note that the second peak around 1.1 eV is very close to the O band (1260-1360 nm) of second telecom window. With photon energy higher than 2.7 eV, there is no higher peak \cite{SM}.

  The calculated half-wave voltage by Eq.~(8) is shown in Fig.~3b, which shows much smaller voltage is required to achieve phase difference of $\pi$ between the two orthogonal optical fields $\frac{1}{\sqrt{2}}|E(\omega)|(1,\pm 1)$. The half-wave voltage can be less than 200 V for visible light. However, the relatively high extinction coefficient $k(\omega)$ of chiral topological semimetal RhSi means rapid attenuation when light propagates in the chiral crystal RhSi. Therefore, the thickness L of electro-optic crystal defined in Eq.~(7) should be as small as possible if the linear electro-optic effect is exploited to modulate the phase velocity of light signal. Similar to graphene based electro-absorption modulator \cite{XZ11}, RhSi can be also used as the metal in metal-oxide-semiconductor capacitor in which the Fermi level of RhSi can be tuned by electric signal. This chiral crystal RhSi based electro-absorption modulator can perform intensity modulation of optical signal. Overall, chiral crystal RhSi can be used as both phase and intensity modulation because of its giant linear electro-optic coefficient and optical absorption.


In summary, we studied the electro-optic effect in chiral topological semimetal RhSi charactized by high-fold chiral fermions separated in energy space. Our results reveals that the general second order conductivity $\sigma^{(2)}_{xyz}$ includes a real symmetric component related to metric tensor and an imaginary antisymmetric component related to curvature tensor, and suggest a giant electro-optic effect in chiral topological semimetal RhSi as large as 7000 pm/V. Especially, the half-wave voltage is in order of hundreds volt for most visible light. This work shows that chiral topological semimetals have potential application in electro-optic modulation with relatively low half-wave voltage for for wavelength in the second telecom window of optical fiber communications. Especially, the half-wave voltage can be further lowered if chiral topological semimetal with higher Chern number is adopted.

This work was supported by the National Natural Science Foundation of China (Grant No. 11604068). Yu Gu is supported by National Key R\&D
Program of China (Grant No. 2017YFA0305500), and Natural Science Foundation of Jiangsu Province (Grant No. BK20200071). T.I. is supported by MEXT via ¡°Exploratory Challenge on Post-K Computer¡± (Frontiers of Basic Science: Challenging the Limits). The calculations were performed on the Hokusai system (Project No. Q21246) of Riken. The authors are grateful to A. V. Pronin for sharing the original data of optical conductivity and helpful discussion.




\begin{thebibliography}{99}
\bibitem{EK21}G. Sinatkas, T. Christopoulos, O. Tsilipakos, and E. E. Kriezis, J. Appl. Phys.  {\bf 130}, 010901 (2021).
\bibitem{Boyd}R. W. Boyd, Nonlinear Optics (Academic, Cambridge, 2003).
\bibitem{QL20}M. Li, J. Ling, Y. He, U. A. Javid, S. Xue, and Q. Lin, Nat. Commun.  {\bf 11}, 4123 (2020).
\bibitem{ML18}C. Wang, M. Zhang, X. Chen, M. Bertrand, A. Shams-Ansari, S. Chandrasekhar, P. Winzer, and M. Lon\v{c}ar, Nature  {\bf 562}, 101-104 (2018).


\bibitem{SB87}R. Soref, and B. Bennett, IEEE Journal of Quantum Electronics  {\bf 23}, 123-129 (1987).
\bibitem{MP04}A. Liu, R. Jones, L. Liao, D. Samara-Rubio, D. Rubin, O. Cohen, R. Nicolaescu, and M. Paniccia, Nature  {\bf 427}, 615-618 (2004).
\bibitem{ML05}Q. Xu, B. Schmidt, S. Pradhan, and M. Lipson, Nature  {\bf 435}, 325-327 (2005).
\bibitem{XX18}C. Haffner, D. Chelladurai, Y. Fedoryshyn, A. Josten, B. Baeuerle, W. Heni, T. Watanabe, T. Cui, B. Cheng, S. Saha, D. L. Elder, L. R. Dalton, A. Boltasseva, V. M. Shalaev, N. Kinsey, and J. Leuthold, Nature  {\bf 556}, 483-486 (2018).
\bibitem{A17}M. Ayata, Y. Fedoryshyn, W. Heni, B. Baeuerle, A. Josten, M. Zahner, U. Koch, Y. Salamin, C. Hoessbacher, C. Haffner, D. L. Elder, L. R. Dalton, J. Leuthold, Science  {\bf 358}, 630-632 (2017).


\bibitem{XZ11}M. Liu, X. Yin, E. Ulin-Avila, B. Geng, T. Zentgraf, L. Ju, F. Wang, and X. Zhang, Nature {\bf 474}, 64-67 (2011).
\bibitem{ML15}C. T. Phare, Y.-H. D. Lee, J. Cardenas, and  M. Lipson, Nature Photon.  {\bf 9}, 511-514(2015).
\bibitem{KH19}K. Chen, X. Zhou, X. Cheng, R. Qiao, Y. Cheng, C. Liu, Y. Xie, W. Yu, F. Yao, Z. Sun, F. Wang, K. Liu, and Z. Liu, Nature Photon.  {\bf 13}, 754-759(2019).

\bibitem{RhSi} S. Geller, and E. A. Wood, Acta Cryst. {\bf 7}, 441-443 (1954).
\bibitem{DH21}B. Q. Lv, T. Qian, and H. Ding, Rev. Mod. Phys.  {\bf 93}, 025002 (2021).
\bibitem{AV18}N. P. Armitage, E. J. Mele, and A. Vishwanath, Rev. Mod. Phys.  {\bf 90}, 015001 (2018).
\bibitem{JO21}J. Orenstein, J. E. Moore, T. Morimoto, D.H. Torchinsky, J.W. Harter, and D. Hsieh, Annual Review of Condensed Matter Physics,  {\bf 12}, 247-272 (2021).

\bibitem{MZH18}G. Chang, B. J. Wieder, F. Schindler, D. S. Sanchez, I. Belopolski, S.-M. Huang, B. Singh, D. Wu, T.-R. Chang, T. Neupert, S.-Y. Xu, H. Lin, and M. Z. Hasan, Nat. Mater. {\bf 17}, 978 (2018).

\bibitem{Sato19}D. Takane, Z. Wang, S. Souma, K. Nakayama, T. Nakamura, H. Oinuma, Y. Nakata, H. Iwasawa, C. Cacho, T. Kim, K. Horiba, H. Kumigashira, T. Takahashi, Y. Ando, and T. Sato, Phys. Rev. Lett. {\bf 122}, 076402 (2019).


\bibitem{MZH17} G. Chang, S.-Y. Xu, B. J. Wieder, D. S. Sanchez, S.-M. Huang, I. Belopolski, T.-R. Chang, S. Zhang, A. Bansil, H. Lin, and M. Z. Hasan, Phys. Rev. Lett. {\bf 119}, 206401 (2017).
\bibitem{ZSC17} P. Tang, Q. Zhou, and S. -C. Zhang, Phys. Rev. Lett. {\bf 119}, 206402 (2017).
\bibitem{DH19}Z. Rao, H. Li, T. Zhang, S. Tian, C. Li, B. Fu, C. Tang, L. Wang, Z. Li, W. Fan, J. Li, Y. Huang, Z. Liu, Y. Long, C. Fang, H. Weng, Y. Shi, H. Lei, Y. Sun, T. Qian, and H. Ding, Nature {\bf 567}, 496 (2019).
\bibitem{MZH19}D. S. Sanchez, I. Belopolski, T. A. Cochran, X. Xu, J.-X. Yin, G. Chang, W. Xie, K. Manna, V. S\"{u}ss, C.-Y. Huang, N. Alidoust, D. Multer, S. S. Zhang, N. Shumiya, X. Wang, G.-Q. Wang, T.-R. Chang, C. Felser, S.-Y. Xu, S. Jia, H. Lin, and M. Z. Hasan, Nature {\bf 567}, 500(2019).
\bibitem{YL19}N. B. M. Schr\"{o}ter, D. Pei, M. G. Vergniory, Y. Sun, K. Manna, F. de Juan, J. A. Krieger, V. S\"{u}{\ss}, M. Schmidt, P. Dudin, B. Bradlyn, T. K. Kim, T. Schmitt, C. Cacho, C. Felser, V. N. Strocov, and Y. Chen, Nat. Phys. {\bf 15}, 759-765 (2019).

\bibitem{CF20}N. B. M. Schr\"{o}ter, S. Stolz, K. Manna, F. de Juan, M. G. Vergniory, J. A. Krieger, D. Pei, T. Schmitt, P. Dudin, T. K. Kim, C. Cacho, B. Bradlyn, H. Borrmann, M. Schmidt, R. Widmer, V. N. Strocov, and C. Felser, Science {\bf 369}, 179-183(2020).
\bibitem{AVP21}L. Z. Maulana, Z. Li, E. Uykur, K. Manna, S. Polatkan, C. Felser, M. Dressel, and A. V. Pronin, Phys. Rev. B  {\bf 103}, 115206 (2021).

\bibitem{JES00}J. E. Sipe and A. I. Shkrebtii, Phys. Rev. B  {\bf 61}, 5337 (2000).
\bibitem{Deyo09} E. Deyo, L. E. Golub, E. L. Ivchenko, and B. Spivak, arXiv:0904.1917.
\bibitem{AC16}A. Cortijo, Phys. Rev. B {\bf 94}, 235123 (2016).
\bibitem{ZL211}Z. Li, S. Zhang, T. Tohyama, X. Song, Y. Gu, T. Iitaka, H. Su, and H. Zeng, Sci. China-Phys. Mech. Astron. {\bf 64}, 107211 (2021).
\bibitem{ZL21B}Y. Gao, T. Iitaka, and Z. Li, Eur. Phys. J. B {\bf 94}, 95 (2021).
\bibitem{ZL19}Z. Li, T. Iitaka, H. Zeng, and H. Su, Phys. Rev. B {\bf 100}, 155201 (2019).

\bibitem{CPGE1} F. de Juan, A. G. Grushin, T. Morimoto, and J. E. Moore, Nat. Commun.  {\bf 8}, 15995 (2017).
\bibitem{NN18} Y. Zhang, H. Ishizuka, J. van den Brink, C. Felser, B. Yan, and N. Nagaosa, Phys. Rev. B {\bf 97}, 241118(R) (2018).

\bibitem{TM18}F. Flicker, F. de Juan, B. Bradlyn, T. Morimoto, M. G. Vergniory, and A. G. Grushin, Phys. Rev. B  {\bf 98}, 155145 (2018).
\bibitem{YS20}C. Le, Y. Zhang, C. Felser, and Y. Sun, Phys. Rev. B  {\bf 102}, 121111(R) (2020).

\bibitem{JO19}D. Rees, K. Manna, B. Lu, T. Morimoto, H. Borrmann, C. Felser, J. E. Moore, D. H. Torchinsky, and J. Orenstein, Sci. Adv. {\bf 6}, eaba0509 (2020).
\bibitem{JE20}A. Avdoshkin, V. Kozii, and J. E. Moore, Phys. Rev. Lett. {\bf 124}, 196603 (2020).
\bibitem{DHT21}D. Rees, B. Lu, Y. Sun, K. Manna, R. \"{O}zg\"{u}r, S. Subedi, H. Borrmann, C. Felser, J. Orenstein, and D. H. Torchinsky, Phys. Rev. Lett. {\bf 127}, 157405 (2021).

\bibitem{TGP17}Alireza Taghizadeh, F. Hipolito, and T. G. Pedersen, Phys. Rev. B {\bf 96}, 195413 (2017).


\bibitem{WML10}Y.-Q. Ma, S. Chen, H. Fan, and W.-M. Liu, Phys. Rev. B {\bf 81},  245129 (2010).
\bibitem{AP17}M. Kolodrubetz, D. Sels, P. Mehta, A. Polkovnikov, Physics Reports {\bf 697},  1-87 (2017).
\bibitem{FP21}A. Graf, and F. Pi\'{e}chon, Phys. Rev. B {\bf 104},  085114 (2021).
\bibitem{DC21}P. Bhalla, K. Das, D. Culcer, A. Agarwal, arXiv:2108.04082.
\bibitem{NN20}J. Ahn, G.-Y. Guo, and N. Nagaosa, Phys. Rev. X {\bf 10},  041041 (2020).

\bibitem{SM}See Supplemental Material at http://link.aps.org/supplemental/XXXXXXXX for detailed derivations of linear and nonlinear optic conductivity with dipole approximation, and calcualted optical conductivity, electro-optic coeffecient in large photon energy scale.
\bibitem{ELK}E. Sj\"{o}stedt, L. Nordstr\"{o}m, and D. J. Singh, Solid State Communications {\bf 114}, 15-20 (2000).
\bibitem{ABINIT}X. Gonze, F. Jollet, F.A. Araujo, D. Adams, B. Amadon, T. Applencourt, C. Audouze, J.-M. Beuken, J. Bieder, A. Bokhanchuk, et al., Comput. Phys. Commun., {\bf 205}, 106-131 (2016).
\bibitem{SOS}S. Sharma, and C. Ambrosch-Draxl, Phys. Scr. {\bf T109}, 128 (2004).


\bibitem{AVP20}L. Z. Maulana, K. Manna, E. Uykur, C. Felser, M. Dressel, and A. V. Pronin, Phys. Rev. Research {\bf 2}, 023018 (2020).
\bibitem{WL20} Z. Ni, B. Xu, M.-\'{A}. S\'{a}nchez-Mart\'{\i}nez, Y. Zhang, K. Manna, C. Bernhard, J. W. F. Venderbos, F. de Juan,
C. Felser, A. G. Grushin, and L. Wu, npj Quantum Materials {\bf 5}, 96 (2020).
\bibitem{Fu15}I. Sodemann and L. Fu, Phys. Rev. Lett. {\bf 115}, 216806 (2015).










































\end{thebibliography}

\end{document}